# Emotions, diffusive emotional control and the motivational problem for autonomous cognitive systems

C. Gros, Department of Physics, J.W. Goethe University Frankfurt, gros07[at]itp.uni-frankurt.de

**Abstract**

All self-active living beings need to solve the motivational problem: The question what to do at any moment of their live. For humans and non-human animals at least two distinct layers of motivational drives are known, the primary needs for survival and the emotional drives leading to a wide range of sophisticated strategies, such as explorative learning and socializing. Part of the emotional layer of drives has universal facets, being beneficial in an extended range of environmental settings. Emotions are triggered in the brain by the release of neuromodulators, which are, at the same time, the agents for meta-learning. This intrinsic relation between emotions, meta-learning and universal action strategies suggests a central importance for emotional control for the design of artificial intelligences and synthetic cognitive systems. An implementation of this concept is proposed in terms of a dense and homogeneous associative network (dHan).

**Introduction**

Is it a coincidence, a caprice of nature, that the species living presently on our planet with the most developed intellectual and cognitive capabilities, humanity, is also thoroughly infused with emotions? Or is it a conditio sine qua non: Are higher cognitive powers intrinsically dependent on a functioning and solid emotional grounding? This question is centrally relevant for our scientific and philosophical self-understanding, posing at the same time a paradigmatic challenge for the development of synthetic cognitive systems and artificial intelligences (AI).

A wide range of different notions are connected with the term emotion and with the personal experience of emotions (Barrett, Mesquita, Ochsner, Gross, 2007). Social interactions and emotional involvements, to give an example, take-up a good share of our daily life and the social aspects of emotional expressions are being widely discussed (Blakermore, Winston & Frith, 2004; Lieberman, 2007). They constitute an important aspect in human-robot interactions (Breazeal, 2003) and may even play a role in human phylogenesis (Parr, Waller & Fugate, 2005), having a high adaptive value (Rolls, 2005). The study of synthetic emotions (Picard, 2000) constitutes therefore a field of growing importance, dealing, beside others, with the role of emotions in artificial intelligences in general (Minsky, 2007), social robots (Duffy, 2003; Fong, Nourbakhsh & Dautenhahn, 2003), emotional expression in speech and language (Murray & Arnott, 2008) and social synthetic computer characters (Tomlinson & Blumberg, 2002).

It is well known, that emotions are triggered by neuromodulators like dopamine, serotonin and opioids, and that the very same neuromodulators can be found all over the animal kingdom, and not just in mammals (Arbib & Fellous, 2004). It is therefore reasonable to assume, that the neurobiological foundations of emotion-like functionalities, being present to a varying extend in all animals having a central or distributed nervous system, precedented phylogenetically higher cognitive capabilities, like sophisticated social interactions or logical reasoning. This observation suggests an underlying functional role of emotions, or emotion-like regulative processes, for both simple and highly developed cognitive systems in general. Neurobiological studies have found indeed close relations between emotions and the internal reward system (Aron *et. Al,* 2005; Kringelbach, 2005; Burgdorf & Panksepp,

2006), indicating that there is a close relation between emotions and decision making (Damasio, 1994; Naqvi, Shiv & Bechara, 2006; Coricelli, Dolan & Sirigu, 2007) quite in general. In the following we will describe, from the functional perspective of dynamical system theory, the role of emotions in cognitive systems. Taking into account the established results from experimental neurobiology and experimental psychology, a theory for emotions will emerge that can be translated algorithmically precisely into formulas and code lines, a prerequisite for the realization of synthetic emotions in artificial intelligences and robots.

**Motivations**

In order to elucidate the general functional purposes of emotions we start by considering the motivational problem of self-determined living creatures, whether biological or artificial. We use here and in the following the general term `cognitive system' for such an autonomous and self-determined being. The question then regards the general motivational drives for cognitive systems.

The basic motivational drive of all living organisms is the `instinct for survival' and it is sometimes assumed, indeed this is the general folklore in the larger public, that the survival instinct would be the sole driving force. In this context the desire to survive would determine in ultima ratio all activities of non-human animals, as well as the ones of humans, e.g. the decision to attend a violin concert instead of a cello performance.

Cognitive systems are instances of complex and adaptive dynamical systems (Gros, 2008) and the survival instinct can be defined algorithmically in a very precise manner, as we will do further below, in terms of a set of survival variables representing the health-status of their respective bodies. Nevertheless, the separate motivational layer, the network of emotions, has several stand-alone features. Emotions might indeed be triggered by the processes representing the survival instinct, but generally they constitute an independent dynamical component. The evolutionary fitness of an animal is increased both by a functioning survival instinct and by a suitable emotional framework (Fellous & Arbib, 2005), but this matter of fact does not imply that both processes have identical causes.

Neuromodulators are the neurobiological roots of emotions (Fellous, 1999) and in the following we will first discuss their biological functionalities in general terms. We will be interested, in particular, in the interplay between local and non-local homeostasis, meta-learning and the diffusive learning signals at basis of the diffusive emotional control. We will find that cognitive systems lacking a diffusive regulative network akin to the one of neuromodulators in the brain, are not likely to have the potential for higher cognitive capabilities. We will then discuss the implications hereof for synthetic cognitive systems in general and then proceed to formulate concrete algorithmical implementations of diffusive emotional control for generalized neural network architectures in the framework of dynamical system theory.

In conclusion, we will find that higher-level cognitive systems lacking diffusive emotional control are not likely to exist, that human-level artificial intelligences based on logical reasoning and the survival instinct alone are probably not possible. We will also see that an algorithmic implementation of diffusive emotional control is possible for synthetic cognitive systems and then shortly discuss that the resulting `true synthetic emotions' will be quite alien to human emotions, as we experience them ourselves.

**Neuromodulators**

Neuromodulators act, from a neurobiological point of view, as a diffusive control system, influencing not the firing state of individual neurons but the responsiveness in general of extended neural ensembles, and even of entire brain regions. From the perspective of dynamical system theory, neuromodulators are therefore the agents for `meta-learning' and homeostasis (Doya, 1999; Marder & Goaillard, 2006), the regulation of slow dynamical variables such as firing thresholds and synaptic sensibility, occurring either automatically or in response to internal or external status signals.

Homeostasis and autoregulation are ubiquitous in biological processes in general, and in the brain in particular (Turrigiano & Nelson 2004). Every individual neuron adapts its average responsiveness, e.g. its firing threshold, relative to the input it receives over time from afferent neurons. This example for a basic local homeostatic process determines the normal or average properties of neurons on an individual basis. The average properties of neurons can be influenced, in addition, by neuromodulators like dopamine, serotonin, and opioids. This regulation of slow variables by neuromodulators is, on the other hand, a process involving several distinct brain structures. Dopamine or serotonin neurons affecting cortical neural ensembles typically receive their signals from subcortical structures, like the amygdala (Phelps, 2006), neuromodulation is intrinsically non-local.

Emotions and neuromodulators are intrinsically linked, but not identical (Damasio, 1994; Fellous, 1999). There are probably no emotions without the concurrent release of neuromodulators, but the brain is a complex and recurrent dynamical system. The geometry of the neuro-chemical information flow is generally not uniquely directed in the brain, feedback loops are ubiquitous. The cognitive information processing and the neuromodulatory component are therefore strongly interacting. Emotional motivation may precede thinking (Balkenius, 1993), but cognitive control of emotions is also possible, and manifestly pronounced in humans (Grey, 2004).

What makes then non-local homeostatic regulation by neuromodulators `emotional', in contrast to the automatic local homeostatic processes occurring on cellular basis, which we may term `neutral'? Introspective experience and a vast body of clinical research data show that emotions and the organization of behavior through motivational drives are intrinsically related (Arbib & Fellous, 2004). When behavior in response to a given emotional arousal is not genetically predetermined, as it is generally the case for highly developed cognitive systems, then the cognitive system needs to learn an adequate response strategy. Algorithmically, this is achieved via reinforcement or temporal-difference learning (Sutton & Barto, 1998). These learning processes avail themselves of reward signals and a given behavioral response will be enhanced or suppressed for positive and negative reward signals respectively. A prominent candidate for a reward signal in the brain is dopamine (Iversena & Iversena, 2007). From this perspective one then concludes, that emotional diffusive control is characterized by a coupling of the regulative event to the generation of reward signals for subsequent reinforcement learning processes.

The key question is then: How are the reward signals generated? Let us consider an example. If we are angry, we will generally try to perform actions with the intent of reducing our level of angriness. When this goal is achieved we then are, usually at least, content. That is, a positive reward signal, reinforcing the precedent behavior, has been generated. Generalizing this example we may formulate the working hypothesis, that the generation of reward signals is coupled to the activation-level of the emotional diffusive regulative control processes. Let us note, that there is at present no direct clinical evidence for the overall validity of this working hypothesis. It is however very powerful, yielding directly a precise prescription for the algorithmic implementation of diffusive emotional control for synthetic cognitive systems and artificial intelligences. Emotional diffusive control then corresponds to regulated meta-learning. The optimal intensity, or the optimal frequency, of a regulated meta-learning process has a genetically preset value and the reinforcement signal is generated when the meta-learning is activated too often or too rarely.

To conclude this section let us return to the initial question, whether a highly developed cognitive system without emotions, viz without non-local homeostatic regulation, is conceivable. The neuromodulators in our brain set our state of mind. Curiosity, anxiety or ebullience, to mention just a few of the myriads of possible emotional states, will generally lead to different behavioral strategies, providing the cognitive system differentiated options for reacting to similar environmental settings. Without the emotional states the cognitive system would be reduced to maximizing its actual survivability probability, or the integrated survivability probability for the foreseeable future. These options however do not constitute an optimal use of resources in environmental situations, to give an example, where surviving is not at stake. A curiosity-driven explorative strategy might then be the better option, potentially increasing the lifetime-fitness of the cognitive system by a substantial amount.

One of the defining characteristics of highly developed cognitive systems is the availability of a wide range of behavioral patterns. Diffusive emotional control provides these capabilities and this road has been taken by evolution, no alternative routes are presently known for synthetic cognitive systems.

## Cognitive Systems

Having discussed the neurobiological functionalities of emotions, we now turn to the case of synthetic cognitive systems. Let us start by considering the defining properties of a cognitive system in general.

Intuitively one may be tempted to identify the human cognitive system with the brain, viz with the physical brain tissue. This is however inappropriate, a cognitive system is strictly speaking an abstract identity, a complex dynamical system consisting of a (very large) set of state variables together with equations determining the time evolution of these variables. The cognitive system takes however `life' only once it becomes embodied, viz when it receives information through appropriate sensors or sensory organs and when it becomes able to perform action through appropriate actuators or limbs. The central defining characteristic of a cognitive system lies in its capability to retain a physical support unit, viz a body, functioning and alive, at least for a certain period of time. This task takes place in a continuously changing environment, as illustrated in Fig. 1. A cognitive system is therefore an instance of what can be termed a `living dynamical system'.

It is interesting to point out in this context, that the physical brain tissue of a person is a part of the environment, and the human cognitive system is the sum of the biophysical processes resulting from the neural brain activity. Philosophical niceties apart, we may define with `environment' everything in the physical world the cognitive system may obtain sensory information about, either directly or indirectly via appropriate instruments. And indeed, we may obtain, at least as a matter of principle, knowledge about the complete physical-chemical state of every one of our own constituting neurons.

## Survival Variables

The primary task of a cognitive system is to keep its own support unit alive. Technically we can define a set of survival variables and the survival instinct then corresponds to the task of keeping these survival variables in a genetically given range. Typical examples for survival variables of biological beings are the blood sugar level, the blood pressure or the heart beating frequency. A classical survival variable for a robotic cognitive system is the battery status. Simple cognitive systems are equipped with preset responses for deviations of the survival variables from their target values, like the simple uptake of food in case of hunger, or the search for a socket when the battery is low. More sophisticated cognitive systems will generally need to acquire adequate responses by learning. E.g. they might need to learn which kinds of food or plant actually reduce the level of hunger and which do not, or how to find the next socket in an artificial labyrinth.

The programming of most real-world robots and AI-programs may be cast into this framework. A chess program typically has just one survival variable, the chance of winning the game. The value of this variable is evaluated via sophisticated deep-search algorithms and the next move it determined by the condition of maximizing the chance of winning the game, viz the probability of survival.

Technically, the implementation of a generalized survival instinct for synthetic cognitive systems does not pose any problem of principle. The actual distance of the survival variables from their given target value can be taken as a measure for the inverse probability of surviving and any action of the system resulting in an increase or in a decrease of the survivability probability will then trigger a positive or a negative reinforcement signal. This reinforcement signal can then be used for appropriate internal supervised learning, increasing or decreasing respectively the probability that the same course of action will be taken in the future for similar environmental conditions. The positioning of the survival instinct within the motivational structure of a cognitive system is illustrated in Fig. 2.

**Autonomous Dynamics**

The simplest conceivable cognitive systems would just react in predetermined ways to incoming sensory stimuli. These responses might be simple, like the flight instinct in the case of danger, or computationally demanding. A soccer-playing robot reacts to the environmental situation, the current position and the velocity of the ball and of the other players, evaluating complex algorithmic routines. The soccer-playing robot is autonomous in the sense that it does not need a human controller. The robots participating in Robo-Cup are however not self-active in the terms of cognitive system theory. At no point does the soccer playing robot consider alternative action strategies; the robot is forced by its programming to continue playing soccer until the game is finished or the battery breaks down. The soccer playing robot will not interrupt playing because of anger or curiosity, it has just one possible `state of mind'. No conflicting internal emotions or states of mind will distract the soccer playing robot.

On a higher level, a cognitive system would dispose of non-trivial internal processes. To classify as autonomous or self-induced, these dynamical processes would need to continue indefinitely even in the absence of sensory stimuli. The internal dynamics remains active even in the presence of a static or quasi-static environment, when nothing is happening in the outside world. One could say, the system is continuously thinking by itself. For mammalian brains this is a well-known and defining neurobiological characteristic. The neural activities of higher cortical areas of mammalian brains are influenced and modulated by sensory stimuli, but not directly driven (Fiser, Chiu & Weliky, 2004). The response is generally not forced. We are hence interested in the interplay of self-generated cognitive activity and emotional control in autonomous cognitive systems.

**Associative Thinking**

We have developed a model system implementing algorithmically the principles of an autonomous cognitive system (Gros, 2005; Gros, 2007). The dHan model (dense Homogenous Associative Network) exhibits self-generated associative thought processes, which we postulate as the driving forces for the self-generated dynamical activities. At any given time only a subset of neurons is active, for a certain period, with the activities of competing neural centers being suppressed. Subsequently a different, in general partially overlapping group of neurons becomes active transiently, such forming an ongoing and never ending series of transient neural activity patterns. This type of neural dynamics, the transient-state dynamics, is illustrated in Fig. 3. For the mathematical formulation implementing these principles we refer to the literature (Gros, 2005; Gros, 2007).

There are findings from experimental neurobiology pointing towards the importance of transient-state dynamics (Abeles et al, 1995; Kenet et al, 2003), indicating that competition and anti-correlation are central organizational principles for the neural activity in the brain (Fox et al, 2005). The transient plateaus in the level of neural activity of a subset of neurons or neural ensembles are also termed `states of the mind' (Edelman & Tononi, 2000) or `winning coalitions'. The composition of the winning coalition changes dynamically from one transient state to the subsequent, giving rise to a vast number of possible states of the mind. The dHan model is therefore an example of a biologically inspired approach to cognitive system theory, seeking to implement known principles of global brain activity, without attempting to reproduce neurobiological details.

**Input Recognition**

A cognitive system continuously receives sensory input containing information about the external environment and about the status of its physical support unit, its body (see Fig. 1). This flow of stimuli competes with the internal, autonomously generated transient-state dynamics. There are then two time series of events, with no a priori connection: The series of subsequently activated winning coalitions generated internally and the flux of sensory stimuli. The sensory input therefore may or may not make a difference. It may or may not influence the internal dynamics, it may or may not influence the composition of the next winning coalition. A primary task of the cognitive system is consequently to find out whether this happens (Gros & Kaczor, 2008). This is a typical task, we term it `input recognition', for diffusive control. We have developed a model, where the interplay between the internal dHan dynamics and the flow of sensory input is regulated through diffusive input recognition (Gros & Kaczor, 2008).

In Fig. 4 the setup of the system is shown. An input layer provides an input data stream to a dHan layer, which is autonomously active. Every site in the dHan layer receives recurrent input from the dHan layer and feed-forward signals from the input layer. Every site can distinguish between these two kinds of inputs and decide which one is the dominant driving signal. A site can therefore decide by itself, through a local process, whether the sensory input had a driving influence in its activation process. In this case a signal is sent to the diffusive control unit responsible for the input recognition, contributing to the activation of this control unit. When the activation level exceeds a certain threshold a diffusive learning signal is released and the links connecting the input layer with the dHan layer are modified in a Hebbian-like fashion. In this way a non-trivial analysis of the input signals is achieved, resulting in an non-linear independent-component analysis (Gros & Kaczor, 2008) and in a mapping of statistically independent objects in the input-data stream to winning coalitions of the dHan layer. For the details we refer to the literature (Gros & Kaczor, 2008).

**Emotional control**

The diffusive control unit responsible for input recognition described above may work either neutrally or emotionally. For the setup illustrated in Fig. 4, made up just of a single input and a single dHan layer, emotional control would be meaningless and the input recognition is neutral, viz there is no preferred activation level. For a full-fledged embedded cognitive system the situation would however be different and the same control unit might acquire emotional character. The system could get `bored' whenever the input recognition would be inactive for a long time (deprivation of sensory signals), or `stressed' whenever it would be continuously active (overloaded with sensory signals). In either case an additional diffusive signal could be released, a reinforcement signal, with the aim of decreasing the probability that similar situations would come up again in the future.

This example, the task of input recognition is a task quite generally necessary for cognitive systems, whichever the respective structural and dynamical organization may be. The mechanisms described here, in the context of the model being investigated, may therefore be generalized and adapted to other approaches and concepts for synthetic cognitive systems.

**Conclusion**

The motivational problem of what to do in one's own life lies at the heart of all living. At a high and philosophical level this fact is reflected by an ongoing and never ending search of humanity, the quest for the meaning of life. On a basic level it implies that all actions of a living being, of a cognitive system, are generated internally, and that a thorough understanding of the decision mechanisms is paramount for an eventually successful realization of artificial cognitive systems. Taking inspiration from neurobiological insights, we have delineated here a layered framework for the motivational drives of an autonomously active biological or synthetic cognitive system.

The overall foundation is given by the survival instinct, algorithmically corresponding to the preprogrammed task of keeping the physical support unit, the body of the cognitive system, functioning and alive. When the basic survival is ensured, emotional control takes over. Emotional control is, in general, functionally independent from the basic need to survive. From the evolutionary point of view the survival instinct is needed to guarantee the short-term survival and emotional control to increase life-time fitness via elaborated behavioral strategies. This separation of time scales is reflected algorithmically, with emotional control being responsible for meta learning, the regulation of slow variables via diffusive reinforcement signals. Importantly, the solution outlined here for the motivational problem can be implemented directly, at least as matter of principles, for artificial cognitive systems and robots, realizing synthetic emotions.

The synthetic emotions generated via diffusive emotional control do not correspond to simulations of emotional expressions, as they are investigated in the context of robot-human communication, but to `true internal emotions', being generated by mechanisms and principles roughly analogous to the emotions present in biological cognitive systems. A correspondence of the qualia of such generated synthetic emotions with the emotions of human or non-human animals is however not to be expected for the foreseeable future.

The mechanisms triggering the release of the neuromodulators conveying emotions in mammals may be either predetermined genetically or acquired culturally. Humans may associate the play of a violin with joy or with distress, or just remain unmoved, there are no marked genetic preferences. This implies that there is an extended layer of culturally acquired motivational drives, as illustrated in Fig. 2, above the survival instinct and above the diffusive emotional control. We believe that a full implementation of this three-layered system of motivational drives is a necessary requirement for the eventual realization of human-level artificial intelligences and cognitive systems and that this goal is to date quite distant from the actual status of research.

**Term Definitions**

COGNITIVE SYSTEM
A cognitive system is an abstract identity, consisting of the set of equations determining the time-evolution of the internal dynamical variables. It needs a *physical support unit* in order to function properly, a datum also denoted as `embedded intelligence'. The primary task for a cognitive system is to retain functionality in certain environments. For this purpose it needs an operational physical support unit for acting and for obtaining sensory information about the environment. The cognitive system remains operational as long as its physical support unit, its body, survives. A cognitive system might be either biological (humans and non-human animals) or synthetic. Non-trivial cognitive systems are capable of learning and of adapting to a changing environment. High-level cognitive systems may show various degrees of intelligence.

AUTONOMOUS COGNITIVE SYSTEM
Cognitive systems are generally autonomous, i.e. self-determined, setting their own goals. This implies that they are not driven, under normal circumstances, by external sensory signals. I.e. an autonomous cognitive system is not forced to perform a specific action by a given sensory stimuli. Autonomy does not exclude the possibility to acquire information from external teachers, given that internal mechanisms allow an autonomous cognitive system to decide whether or not to focus attention on external teaching signals. In terms of a *living dynamical system* an autonomous cognitive system possesses a non-trivial and self-sustained dynamics, viz an ongoing autonomous dynamical activity.

## BIOLOGICALLY INSPIRED COGNITIVE SYSTEM

In principle one may attempt to develop artificial cognitive systems starting with an empty blueprint. Biological cognitive systems are at present however the only existing real-world *autonomous cognitive systems* we know of, and it makes sense to make good use of the general insights obtained by neurobiology for the outline of cognitive system theory. An example such a paradigmal insight is the importance of competitive neural dynamics, viz of neural ensembles competing with each other trying to form winning coalitions of brain regions, suppressing transiently the activity of other neural ensembles. Another example is the intrinsic connection between *diffusive emotional control* and learning mechanisms involving *reinforcement signals*.

## UNIVERSAL COGNITIVE SYSTEM

Simple cognitive systems are mostly ruled by preset stimuli-reaction rules. E.g. an earthworm will automatically try to meander towards darkness. Universal principles, i.e. algorithms applicable to a wide range of different environmental settings, become however predominant in highly developed cognitive systems. We humans, to give an example, are constantly, and most of the time unconsciously trying to predict the outcome of actions and movements taking place in the world around us, even if these outcomes are not directly relevant for our intentions at the given time, allowing us to extract regularities in the observed processes for possible later use. Technically this attitude corresponds to a time-series prediction-task which is quite universal in its applicability. We use it, e.g., to obtain unconsciously knowledge on the ways a soccer ball rolls and flies as well as to extract from the sentences we listen-to the underlying grammatical rules of our mother-tongue.

## PHYSICAL SUPPORT UNIT

Also denoted `body' for biological cognitive systems. Generally it can be subdivided into four functional distinct components. (A) The component responsible for evaluating the time-evolution equations of the cognitive system, viz the brain. (B) The actuators, viz the limbs, responsible for processing the output-signals of the cognitive system. (C) The sensory organs providing appropriate input information on both the external environment and on the current status of the physical support unit. (D) The modules responsible for keeping the other components alive, viz the internal organs. Artificial cognitive systems dispose of equivalent functional components.

## META LEARNING

Meta learning and `homeostatic self-regulation' are closely related. Both are needed for the long-term stability of the cognitive system, regulating internal thresholds, learning-rates, attention fields and so on. They do not affect directly the primary cognitive information processing, e.g. they do not change directly the firing state of individual neurons, nor do they affect the primary learning, i.e. changes of synaptic strengths. The regulation of the sensibility of the synaptic plasticities with respect to the pre- and to the post-synaptic firing state is, on the other hand, a prime task for both meta learning and homeostatic self-regulation. Homeostatic self-regulation is local, always active and present, irrespectively of any global signal. Meta learning is, on the other hand, triggered by global signals, the *diffusive control* signals, generated by the cognitive system itself through distinct sub-components.

## DIFFUSIVE CONTROL

Diffusive control is intrinsically related for biological cognitive systems to the release of neuromodulators. Neuromodulators are generally released in the inter-neural medium, from where they physically diffuse, affecting a large ensemble of surrounding neurons. The neuromodulators do not affect directly the cognitive information processing, viz the dynamical state of individual neurons. They act as the prime agents for transmitting extended signals for *meta learning*. Diffusive control signals come in two versions, neutral and emotional. (A) Neutral diffusive control is automatically activated when certain conditions are present in the cognitive system, irrespectively of the frequency and the level of past activations of the diffusive control. (B) Emotional diffusive control has a preset preferred level of activation frequency and strength. Deviation of the preset activity-level results in negative *reinforcement signals,* viz the system feels `uneasy' or `uncomfortable'.

REINFORCEMENT SIGNAL
Reinforcement signals can be either positive or negative, i.e. a form of reward or punishment. The positive or negative consequences of an action, or of a series of consecutive actions, are taken to reinforce or to suppress the likelihood of selecting the same set of actions when confronted with a similar problem-setting in the future. A reinforcement signal can be generated by a cognitive system only when a nominal target outcome is known. When this target value is given `by hand' from the outside, viz by an external teacher, one speaks of `supervised learning'. When the target value is generated internally one speaks of `unsupervised learning'. The internal generation of meaningful target values constitutes the core of the *motivational problem*.

MOTIVATIONAL PROBLEM
Biological cognitive systems are `autonomous', viz they decided by themselves what to do. Highly developed cognitive systems, like the one of mammals, regularly respond to sensory stimuli and information but are generally not driven by the incoming sensory information, i.e. the sensory information does not force them to any specific action. The motivational problem then deals with the central issue of how a highly developed cognitive system selects its actions and targets. This is the domain of instincts and emotions, even for humans. Note, that rational selection of a primary target is impossible, rational and logical reasoning being useful only for the pursue of primary targets set by the underlying emotional network. Most traditional research in artificial intelligence disregards the motivational problem, assuming internal primary goal selection is non-essential and that explicit primary target selection by supervising humans is both convenient and sufficient.

DYNAMICAL SYSTEM
A dynamical system is a set of variables together with a set of rules determining the time-development of theses variables. The time might be either discrete, viz 1,2,3,... or continuous. In the latter case the dynamical system is governed by a set of differential equations. Dynamical system theory is at the heart of all natural laws, famous examples being Newton's law of classical mechanics, the Schrödinger equation of quantum mechanics and Einstein's geometric theory of gravity, general relativity.

LIVING DYNAMICAL SYSTEM
A living dynamical system is a dynamical system containing a set of variables denoted `survival variables'. The system is defined to be living as long as the value of these variables remain inside a certain preset range and defined to be dead otherwise. Cognitive systems are instances of living dynamical systems and the survival variables correspond for the case of a biological cognitive system to the heart frequency, the blood pressure, the blood sugar level and so on.

COMPLEX SYSTEM THEORY
Complex system theory deals with `complex' dynamical systems, viz with dynamical systems containing a very large number of interacting dynamical variables. Preeminent examples of complex systems are the gen-regulation network at basis of all living, self-organizing phase transitions in physics like superconductivity and magnetism, and cognitive systems, the later being the most sophisticated and probably also the least understood of all complex dynamical systems.

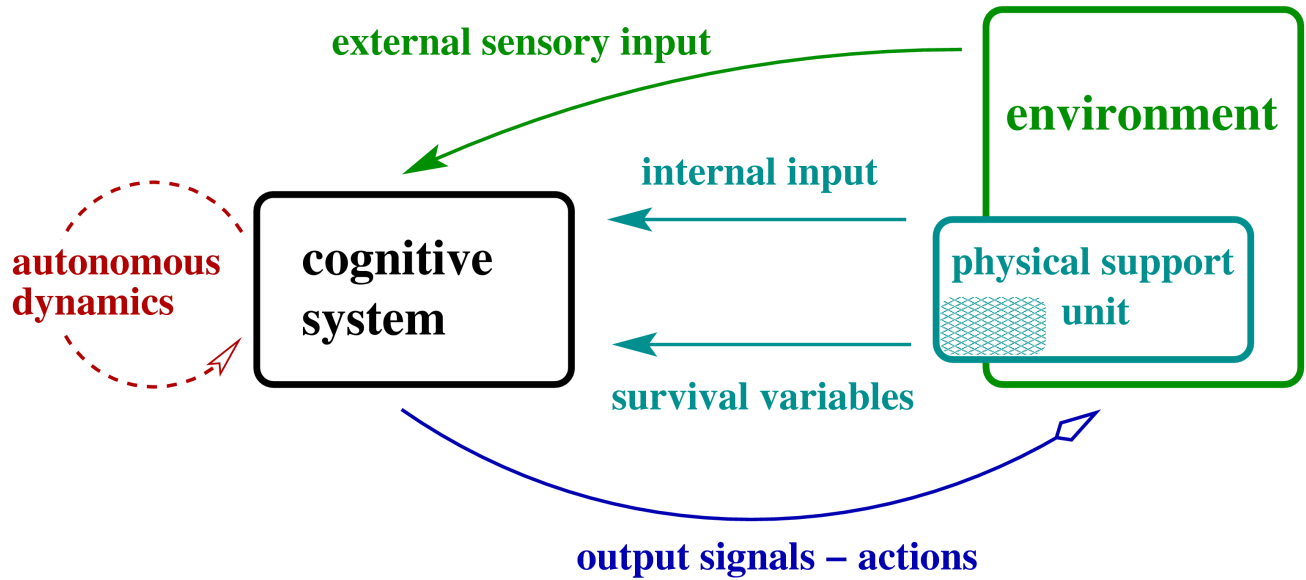

**Fig. 1 – Cognitive System.** Schematic illustration of the interplay between an autonomous, i.e. a self-determined cognitive system and its environment. The cognitive system is an abstract living dynamical system, its time-evolution equations being executed by part of its support unit (shaded region), which corresponds to the brain for a biological cognitive system. Note that its physical support unit, viz its body, is part of the environment from which the cognitive system receives both external and internal sensory input data.

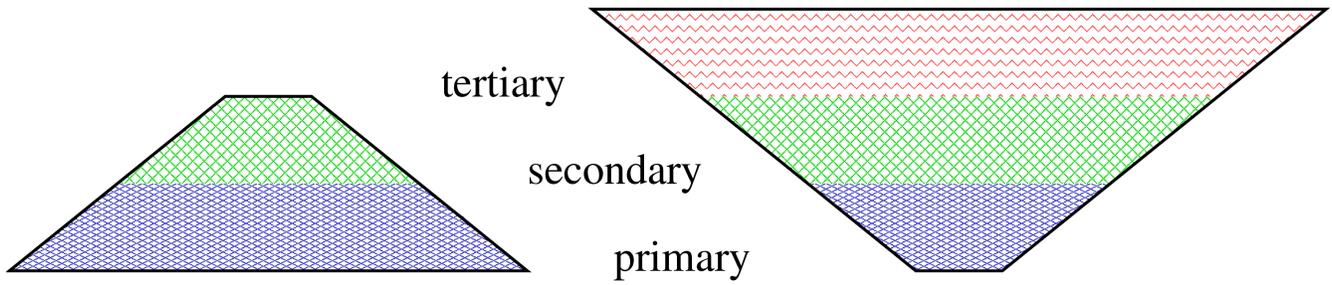

**Fig. 2 – Motivational Pyramids.** Schematic illustration of the motivational pyramids for simple (left drawing) and highly developed (right drawing) biological or synthetic cognitive systems. The primary drives correspond to the genetically encoded survival mechanisms, guaranteeing the basic functionality of the support unit. The secondary drives correspond to the diffusive emotional control setting longer-term goals and survival strategies. The tertiary level correspond to the culturally acquired motivations. Note the predominance of the secondary and the tertiary drives for highly developed cognitive systems.

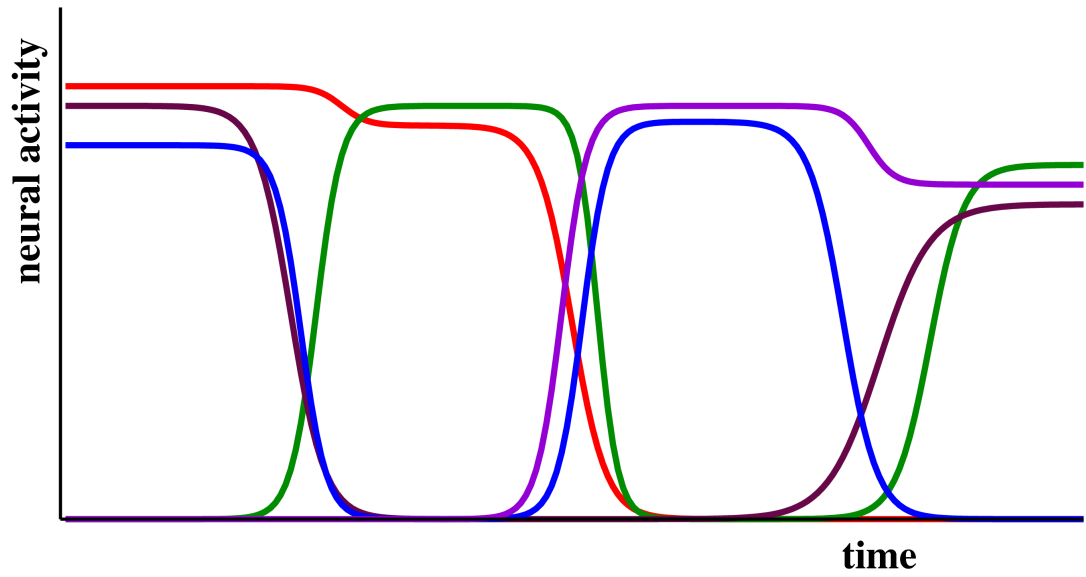

**Fig. 3 – Transient States.** Schematic illustration of a sequence of transiently stable winning coalitions of a neural ensemble. The firing state of any given neuron is either close to zero or transiently stable for a finite period of time, with relatively short transition periods.

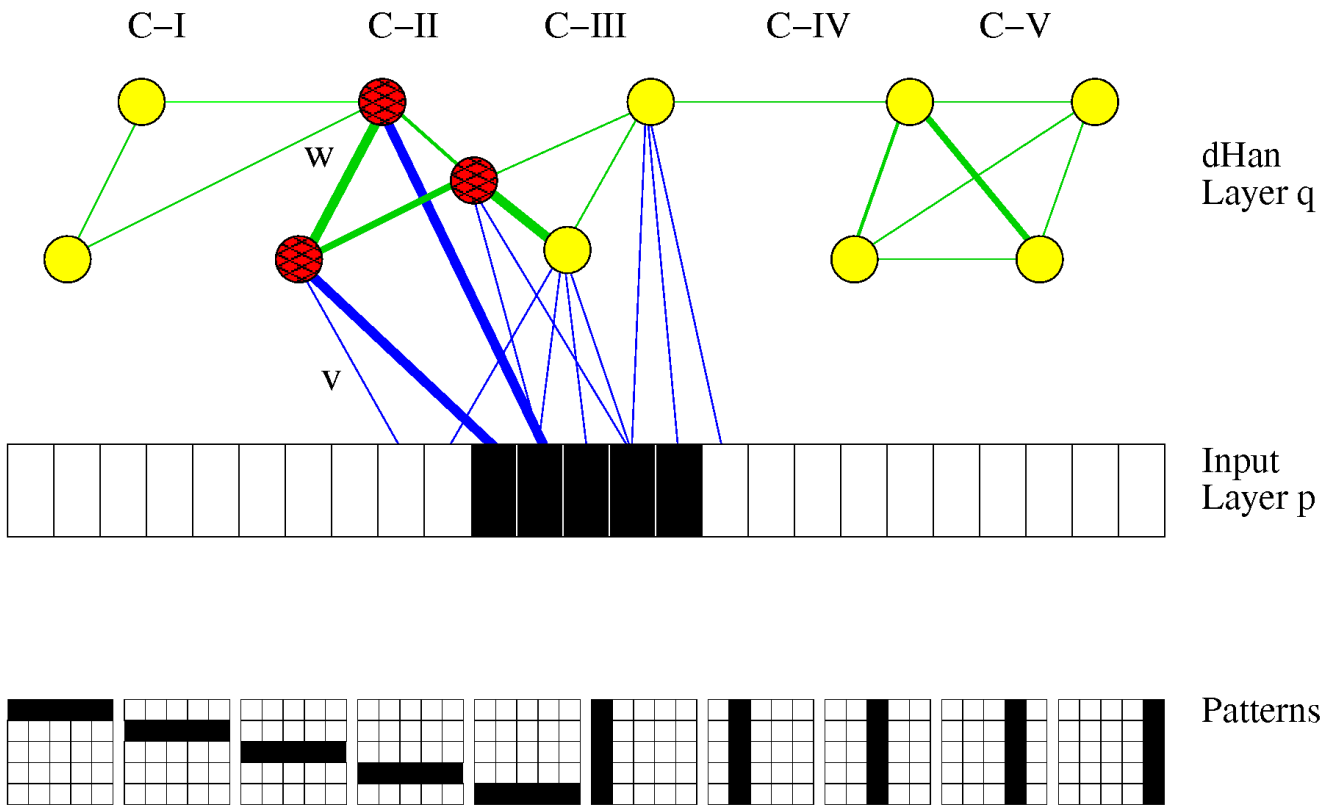

**Fig. 4 – Input Processing.** The model system consisting of a dHan (dense and homogeneous associative network) and an input layer. The input signals are illustrated as raw horizontal and vertical bars. The dHan layer is autonomously active, C-I, ..., C-V denoting the possible winning coalitions of sites. The input signal competes with the internal activity of the dHan layer. The interconnections input-dHan layer are modified during learning, which is activated through an autonomously generated diffusive learning signal.